# Structures of the Molecular Components in DNA and RNA with Bond Lengths Interpreted as Sums of Atomic Covalent Radii

Raji Heyrovská*

Institute of Biophysics of the Academy of Sciences of the Czech Republic, Královopolská 135, 61265 Brno, Czech Republic.

*E-mail: rheyrovs@hotmail.com

**Abstract:**

The author has found recently that the lengths of chemical bonds are sums of the covalent and or ionic radii of the relevant atoms constituting the bonds, whether they are completely or partially covalent or ionic. This finding has been tested here for the skeletal bond lengths in the molecular constituents of nucleic acids, adenine, thymine, guanine, cytosine, uracil, ribose, deoxyribose and phosphoric acid. On collecting the existing data and comparing them graphically with the sums of the appropriate covalent radii of C, N, O, H and P, it is found that there is a linear dependence with unit slope and zero intercept. This shows that the bond lengths in the above molecules can be interpreted as sums of the relevant atomic covalent radii. Based on this, the author has presented here (for the first time) the atomic structures of the above molecules and of DNA and RNA nucleotides with Watson-Crick base pairs, which satisfy the known dimensions.

## INTRODUCTION

Following the exciting discovery [1] of the molecular structure of nucleic acids, the question of the skeletal bond lengths in the molecules constituting DNA and RNA has continued to be of interest. The author was enthused to undertake this work by the recent findings [2a, b] that the length of the chemical bond between any two atoms or ions is, in general, the sum of the radii of the atoms and or ions constituting the bond. The additivity of radii was also found to hold for the hydration bonds [3a] with ions in aqueous solutions as well as for the lengths of the hydrogen bonds [3b] in the Watson-Crick [1b] base pairs of nucleic acids and in many other inorganic and biological compounds.

Pauling [4] treated covalent bond lengths as sums of the covalent radii of the atoms. Although he dealt with the bond lengths in the purines: adenine (A) and guanine (G) and the pyrimidines: thymine (T), cytosine (C) and uracil (U), he did not interpret these bonds. This article shows here that the existing bond lengths [4-7] (see Tabs. 1 and 2) in the above molecules and in ribose (Ri), deoxyribose (De) and phosphoric acid (Ph), correspond to the sums of the appropriate covalent radii of the five atoms, carbon, nitrogen, hydrogen, oxygen and phosphorus. These structures have been combined together as DNA and RNA nucleotides with Watson-Crick base pairs, which satisfy the known crystallographic parameters.

## ATOMIC COVALENT RADII AND THE CN BOND LENGTH IN ADENINE

It was noticed by Pauling [4] that in all the bases, A, T, C, G and U, (see Fig. 1), the CC bond length is around 1.40 Å and the CN bond length has values around 1.32 Å and 1.36 Å. A survey of the existing bond length data [4-7] (see Tabs. 1 and 2), shows that the CC, CN CH, NH, CO, OH and PO bonds in the above molecules do have some essentially fixed bond lengths, and in fact more than the three considered by Pauling.

For the interpretation here of these bond lengths, the atomic covalent radii, defined [4] as $R_{cov} = d(A) = d(AA)/2$, where d(AA) is the covalent bond length between two atoms (A) of the same kind, have been used. Since $R_{cov}$ is half the inter-atomic distance, it is actually a distance, d(A), although denoted by R (for radius). Considering carbon (see Fig. 2A), Pauling [4] points out two types of single bond radii for C: the aliphatic single bond covalent radius of 0.77 Å (this is half the CC tetrahedral bond distance in diamond), and the aromatic single bond (involved in resonance) covalent radius of 0.72 Å (this is half the CC bond distance in the hexagonal plane of graphite). The covalent double bond radius of C, defined [4] as half the C=C double bond length, is 0.67 Å. In general, the covalent double bond radii of atoms are less than those for single bonds and the covalent triple bond radii are less than those for the bond [4].

The above three C (with the subscripts, s.b.: single bond, res.: aromatic resonance bond and d.b.: double bond) are represented by circles of radii $R_{cov}$ in Fig. 2B. Similarly in Fig. 2B, $R_{cov}$ for $N_{s.b.}$ & $N_{d.b.}$ and $O_{s.b.}$ & $O_{d.b.}$ are halves of the single bond and double bond lengths of N and O respectively, $R_{cov}$ for H is half the HH single bond distance [4] and $R_{cov}$ for P was obtained by subtracting $R_{cov}$ of O(I) from the PO(I) bond length [4-7]. Note that the radius for $C_{res}$, 0.72 is the mean of (0.77 + 0.67)/2 the radii for $C_{s.b.}$ and $C_{d.b.}$, as if $C_{res}$ is the radius of a 1.5 bond! In the case of the radius of $N_{d.b}$, 0.62, it is the mean of those of $N_{s.b.}$, 0.70 and $N_{t.b}$, 0.55 [4], where the subscript t.b. denotes triple bond.

An interesting observation from the bond length data in [4-7] assembled in Tab. 1 is e.g., that the CN bond lengths in the aromatic ring in adenine (see col. 2, Tab. 1) are the same [1.34 (+/-) 0.02Å] irrespective of whether they pertain to single (N1-C2 and N3-C4) or double (C2-N3 and C6-N1) bonds. This is similar to the finding in [4] that in the case of benzene, all the six CC bonds are of length 1.39 (+/-) 0.01 Å (with equal bond order [4] of 1.5, due to resonance), although represented by three alternating single and double bonds. This CC bond length of 1.39 Å was interpreted in [2a] as the sum of the radii for $C_{res.}$ ( = 0.72 Å) and $C_{d.b.}$ (= 0.67 Å), see Fig. 2A. Similarly, in adenine, the CN bond distance of 1.34 Å is the sum of $R_{cov}$ of $C_{res.}$ (= 0.72 Å) and $N_{d.b.}$ (= 0.62 Å).

## BOND LENGTHS IN ALL THE MOLECULES AS SUMS OF ATOMIC COVALENT RADII

The above additivity of atomic radii was then tested for the distances between any two atoms in the skeletal structures of the molecular components of nucleic acids by comparing the data from [4-7] with the sums R(sum) of the appropriate atomic covalent radii in Fig. 2B. The data for A, G, T and C are given in Tab. 1 and for U, Ri, De and Ph in Tab. 2. The average values of the bond lengths [4-7] for the above molecules (from Tabs. 1 and 2) and the corresponding R(sum) for the various bonds are tabulated in Tab. 3. On plotting these bond lengths versus R(sum) as shown in Fig. 3, and drawing a least square straight line through all the data points, it was found to have a slope of 1.01 and an intercept of -0.02. The calculated bond lengths using this slope and intercept, (given in the last column in Tab. 3) are found to be identical with R(sum) (see col. 2, Tab. 3). *This shows that the bond lengths in all the above molecules can be considered as the sums of the relevant atomic covalent radii.* Note also that since the straight line in Fig. 3 includes the theoretical values of bond lengths from [6], the above result implies that the latter values are good representations of the radii sum, R(sum).

The above finding has thus enabled the conventional molecular structures (see Fig. 1) to be resolved (for the first time) into the atomic structures based on the individual covalent radii of the constituent atoms. Fig. 4 shows the atomic structures of the bases: T (U), A, C and G and Fig. 5, those for phosphoric acid, deoxyribose (and ribose). For these molecules, while confirming Pauling's [4] two CN bonds of lengths 1.32 and 1.36 (+/- 0.02 Å) and one CC bond length of 1.40 Å, it is found here that actually there are five CN bonds, three CC bonds, two CO bonds and two PO bonds of lengths as shown in Tab. 3.

The bond lengths of C, O and N with H ($R_{cov}$ = 0.37 Å, see Fig. 2B) are also given in Tab. 3. Note that whereas the increase in the CH bond length in going from aromatic to aliphatic C is attributed by Pauling [4] to the change in the radius of H, here it is shown to be due to those of $C_{res.}$ and $C_{s.b.}$.

Fig. 6 shows the atomic structure of the DNA (and RNA) nucleotides with the hydrogen bonded Watson-Crick base pairs (T, A and C, G), in accordance with the known crystallographic parameters in [1c]. The distance between two phosphorus atoms (in pink) which are attached to O(5') and O(3') of a sugar molecule are 7.1 Å and each phosphorus atom is at a distance of 10 Å from the central axis of the helix. The length of 17 Å contains five P atoms in accordance with the length per turn 34 Å of the helix, which contains 10 P atoms. The region of 3.4 Å is a gap between two steps consisting of the T, A and C, G base pairs (which are not in the plane of the paper). A description of the structure is given in the legend for Fig. 6.

Thus, the skeletal bond lengths in the conventional molecular structures of the components and of the nucleotides of DNA and RNA are shown here to be sums of the covalent radii of the two adjacent atoms constituting the bonds (see Figs. 4 - 6).

ACKNOWLEDGEMENTS

The author is grateful to Prof. E. Palecek of the Institute of Biophysics for the moral support, and for the grants AVOZ50040507 of the Academy of Sciences of the Czech Republic and LC06035 of the Ministry of Education, Youth and Sports of the Czech Republic.

**Table 1. Bond lengths (Å) (+/- 0.02Å) [4-7] of adenine (A), guanine (G), thymine (T) and cytosine (C), compared with sums of atomic covalent radii, R(sum) (upper values, in bold; see also Fig. 2B, Tab. 3 and Fig. 4). The numberings of the atoms are as in [7].**

| Adenine | Bond lengths | Guanine | Bond lengths | Thymine | Bond lengths | Cytosine | Bond lengths |
|---|---|---|---|---|---|---|---|
| N1-C2 | **1.34 = 0.62 + 0.72** | N1-C2 | **1.37 = 0.67 + 0.70** | N1-C2 | **1.37 = 0.67 + 0.70** | N1-C2 | **1.37 = 0.67 + 0.70** |
| * | 1.34,1.34,1.32,1.34 | | 1.32,1.37,1.34 | | 1.38,1.38,1.36 | | 1.41,1.40,1.36 |
| C2-N3 | **1.34 = 0.62 + 0.72** | C2-N3 | **1.29 = 0.67 + 0.62** | C2-N3 | **1.37 = 0.67 + 0.70** | C2-N3 | **1.37 = 0.67 + 0.70** |
| * | 1.34,1.33,1.32,1.34 | | 1.36,1.32,1.32 | | 1.38,1.37,1.36 | | 1.37,1.35,1.34 |
| N3-C4 | **1.34 = 0.62 + 0.72** | N3-C4 | **1.29 = 0.67 + 0.62** | N3-C4 | **1.37 = 0.67 + 0.70** | N3-C4 | **1.29 = 0.67 + 0.62** |
| * | 1.34,1.34,1.36,1.34 | | 1.42,1.35,1.32 | | 1.40,1.38,1.36 | | 1.31,1.34,1.32 |
| C4-C5 | **1.39 = 0.67 + 0.72** | C4-C5 | **1.39 = 0.67 + 0.72** | C4-C5 | **1.39 = 0.67 + 0.72** | C4-C5 | **1.39 = 0.67 + 0.72** |
| * | 1.40,1.38,1.40,1.44 | | 1.44,1.38,1.38 | | 1.46,1.45,1.40 | | 1.44,1.43,1.40 |
| C5-C6 | **1.39 = 0.67 + 0.72** | C5-C6 | **1.39 = 0.67 + 0.72** | C5-C6 | **1.39 = 0.67 + 0.72** | C5-C6 | **1.39 = 0.67 + 0.72** |
| * | 1.39,1.41,1.40,1.38 | | 1.39,1.42,1.40 | | 1.35,1.34,1.40 | | 1.36,1.34,1.40 |
| C6-N1 | **1.34 = 0.62 + 0.72** | C6-N1 | **1.37 = 0.67 + 0.70** | C6-N1 | **1.37 = 0.67 + 0.70** | C6-N1 | **1.37 = 0.67 + 0.70** |
| * | 1.34,1.35,1.32,1.34 | | 1.36,1.39,1.36 | | 1.37,1.38,1.36 | | 1.35,1.37,1.34 |
| C5-N7 | **1.29 = 0.67 + 0.62** | C5-N7 | **1.34 = 0.62 + 0.72** | C2-O2 | **1.27 = 0.67 + 0.60** | C2-O2 | **1.27 = 0.67 + 0.60** |
| * | 1.37,1.39,1.32,1.32 | | 1.37,1.39,1.32 | | 1.22,1.22,1.23 | | 1.22,1.24,1.23 |
| N7-C8 | **1.29 = 0.67 + 0.62** | N7-C8 | **1.29 = 0.67 + 0.62** | C4-O4 | **1.27 = 0.67 + 0.60** | C4-N4 | **1.37 = 0.67 + 0.70** |
| * | 1.32,1.31,1.32,1.28 | | 1.32,1.31,1.32 | | 1.22,1.23,1.23 | | 1.36,1.34,1.35 |
| C8-N9 | **1.37 = 0.67 + 0.70** | C8-N9 | **1.37 = 0.67 + 0.70** | C5-M5 | **1.49 = 0.72 + 0.77** | N1-C1' | **1.47 = 0.70 + 0.77** |
| * | 1.37,1.37,1.33,1.37 | | 1.37,1.37,1.32 | | 1.49,1.50,1.53 | | -,1.47,1.53 |
| N9-C4 | **1.42 = 0.70 + 0.72** | N9-C4 | **1.37 = 0.67 + 0.70** | N1-C1' | **1.47 = 0.70 + 0.77** | | |
| * | 1.38,1.37,1.36, - | | 1.37,1.38,1.34 | | -,1.47,1.53 | | |
| C6-N6 | **1.42 = 0.70 + 0.72** | C2-N2 | **1.37 = 0.67 + 0.70** | | | | |
| * | -,1.34,1.35,1.34 | | 1.38,1.34,1.35 | | | | |
| N9-C1' | **1.47 = 0.70 + 0.77** | C6-O6 | **1.27 = 0.67 + 0.60** | | | | |
| * | -,1.46,1.53,- | | 1.22,1.24,1.23 | | | | |
| | | N9-C1' | **1.47 = 0.70 + 0.77** | | | | |
| * | | | -,1.46,1.53 | | | | |
| *Values: | | *Values: | | *Values: | | *Values: | |
| Refs. 6,7,4,5 | | Refs.6,7,4 | | Refs.6,7,4 | | Refs.6,7,4 | |

**Table 2. Bond lengths (+/- 0.02Å) [4-7] of uracil (U), ribose (Ri), deoxyribose (De), phosphoric acid/phosphate (Ph), compared with sums of atomic covalent radii, R(sum) (upper values, in bold, see also Fig. 2B, Tab. 3 and Figs. 4 and 5). The numberings of the atoms are as in [7].**

| Uracil | Bond length | Ribose, de-Oxyribose | Bond length | **Phosphoric acid, phosphate | Bond length |
|---|---|---|---|---|---|
| N1-C2 | **1.37 = 0.67 + 0.70** | C1'-C2' | **1.54 = 0.77 + 0.77** | **P=O | **1.52 = 0.92 + 0.60** |
| * | 1.38,1.34 | | 1.53,1.52 | | 1.49,1.52 |
| C2-N3 | **1.37 = 0.67 + 0.70** | C2'-C3' | **1.54 = 0.77 + 0.77** | **P-O(H) | **1.59 = 0.92 + 0.67** |
| * | 1.37,1.38 | | 1.53,1.52 | | 1.61,1.57 |
| N3-C4 | **1.37 = 0.67 + 0.70** | C3'-C4' | **1.54 = 0.77 + 0.77** | P-O3' | **1.59 = 0.92 + 0.67** |
| * | 1.38,1.37 | | 1.52,1.53 | | 1.61,1.56 |
| C4-C5 | **1.39 = 0.67 + 0.72** | C4'-O4' | **1.44 = 0.77 + 0.67** | P-O5' | **1.59 = 0.92 + 0.67** |
| * | 1.43,1.41 | | 1.45,1.45 | | 1.59,1.56 |
| C5-C6 | **1.39 = 0.67 + 0.72** | O4'-C1' | **1.44 = 0.77 + 0.67** | (P)O5'-C5' | **1.44 = 0.77 + 0.67** |
| * | 1.34,1.41 | | 1.41,1.42 | | 1.44,- |
| C6-N1 | **1.37 = 0.67 + 0.70** | C3'-O3' | **1.44 = 0.77 + 0.67** | (P)O3'-C3' | **1.44 = 0.77 + 0.67** |
| * | 1.38,1.34 | | 1.42,1.43 | | 1.43,- |
| C2-O2 | **1.27 = 0.67 + 0.60** | C5'-C4' | **1.54 = 0.77 + 0.77** | | |
| * | 1.22,1.23 | | 1.51,1.51 | | |
| C4-O4 | **1.27 = 0.67 + 0.60** | C2'-O2' | **1.44 = 0.77 + 0.67** | | |
| * | 1.23,1.24 | | 1.41,- | | |
| N1-C1' | **1.47 = 0.70 + 0.77** | C1'-N1/N9 | **1.47 = 0.70 + 0.77** | | |
| * | 1.47,- | | 1.47,1.47 | | |
| | | O5'-C5' | **1.44 = 0.77 + 0.67** | | |
| * | | | 1.42,1.42 | | |
| *Values: | | *Values: ribose,deoxyribose | | *Values: | |
| Refs. 7,4 | | Ref. 7 | | Refs. 7,4 | |

**Table 3. Average bond lengths (+/- 0.03 Å) [4-7] corresponding to sums of atomic covalent radii, R(sum). Last column: calc. values as per the least sq. line in Fig. 3.**

| | | **Average bond lengths from Tabs. 1 and 2** | | | | | | | **Calculated** |
|---|---|---|---|---|---|---|---|---|---|
| **Bonds** | **R(sum)** | **A** | **G** | **T** | **C** | **U** | **Ri,De** | **Ph** | **bond length** |
| $C_{d.b.}$ - $O_{d.b.}$ | **1.27** | | 1.24 | 1.23 | 1.23 | 1.23 | | | **1.27** |
| $C_{d.b.}$ - $N_{d.b.}$ | **1.29** | 1.33 | 1.32 | | 1.32 | | | | **1.29** |
| $C_{res.}$ - $N_{d.b.}$ | **1.34** | 1.34 | 1.37 | | | | | | **1.34** |
| $C_{d.b.}$ - $N_{s.b.}$ | **1.37** | 1.36 | 1.36 | 1.37 | 1.36 | 1.37 | | | **1.37** |
| $C_{res.}$ - $C_{d.b.}$ | **1.39** | 1.4 | 1.39 | 1.4 | 1.44 | 1.4 | | | **1.39** |
| $C_{res.}$ - $N_{s.b.}$ | **1.42** | 1.35 | | | | | | | **1.42** |
| $C_{s.b.}$ - $O_{s.b.}$ | **1.44** | | | | | | 1.42 | 1.44 | **1.44** |
| $C_{s.b.}$ - $N_{s.b.}$ | **1.47** | 1.5 | 1.5 | 1.45 | 1.5 | 1.47 | 1.47 | | **1.47** |
| $C_{s.b.}$ - $C_{res.}$ | **1.49** | | | 1.48 | | | | | **1.49** |
| P - $O_{d.b.}$ | **1.52** | | | | | | | 1.51 | **1.52** |
| $C_{s.b.}$ - $C_{s.b.}$ | **1.54** | | | | | | 1.52 | | **1.54** |
| P - $O_{s.b.}$ | **1.59** | | | | | | | 1.58 | **1.59** |
| $C_{d.b.}$ – H | **1.04** (A,T,U,C,G) | | | | | | | | |
| $C_{res.}$ – H | **1.09** (A,U,C) | | | | | | | | |
| $C_{s.b.}$ – H | **1.14** (T,Ri,De) | | | | | | | | |
| $N_{s.b.}$ – H | **1.07** (A,T,U,C,G) | | | | | | | | |
| $O_{s.b.}$ – H | **1.04** (Ri,De,Ph) | | | | | | | | |

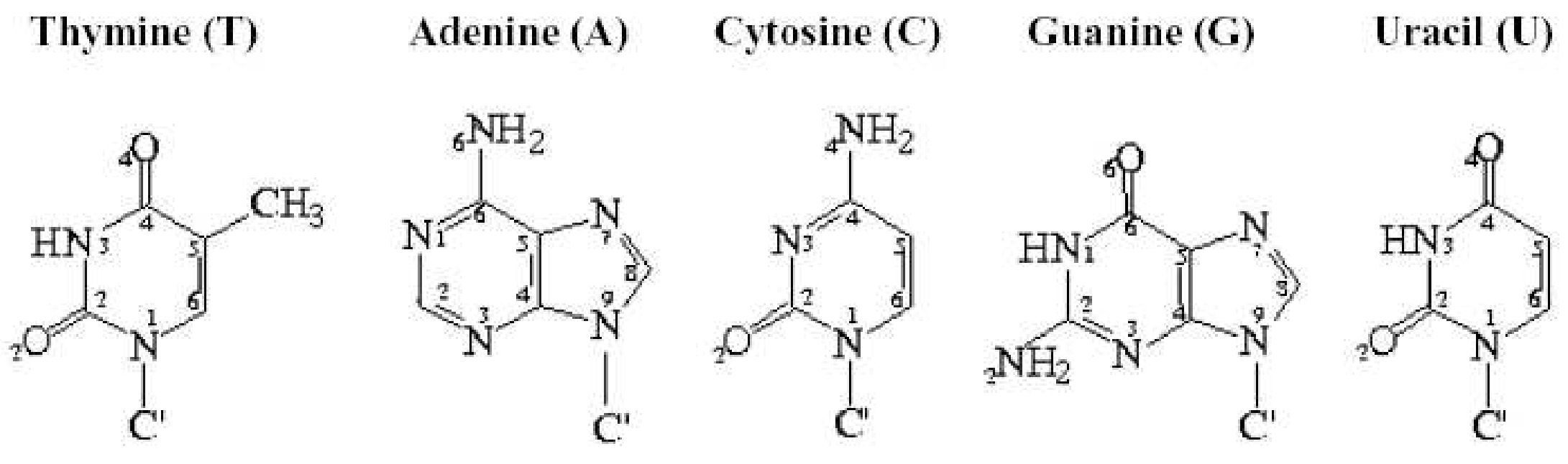

**Figure 1.** Molecular structures [7] of thymine (T), adenine (A), cytosine (C), guanine (G) and uracil (U). Bond length data [4-7] in Tables 1 and 2 correspond to distances between two adjacent (numbered) atoms.

.

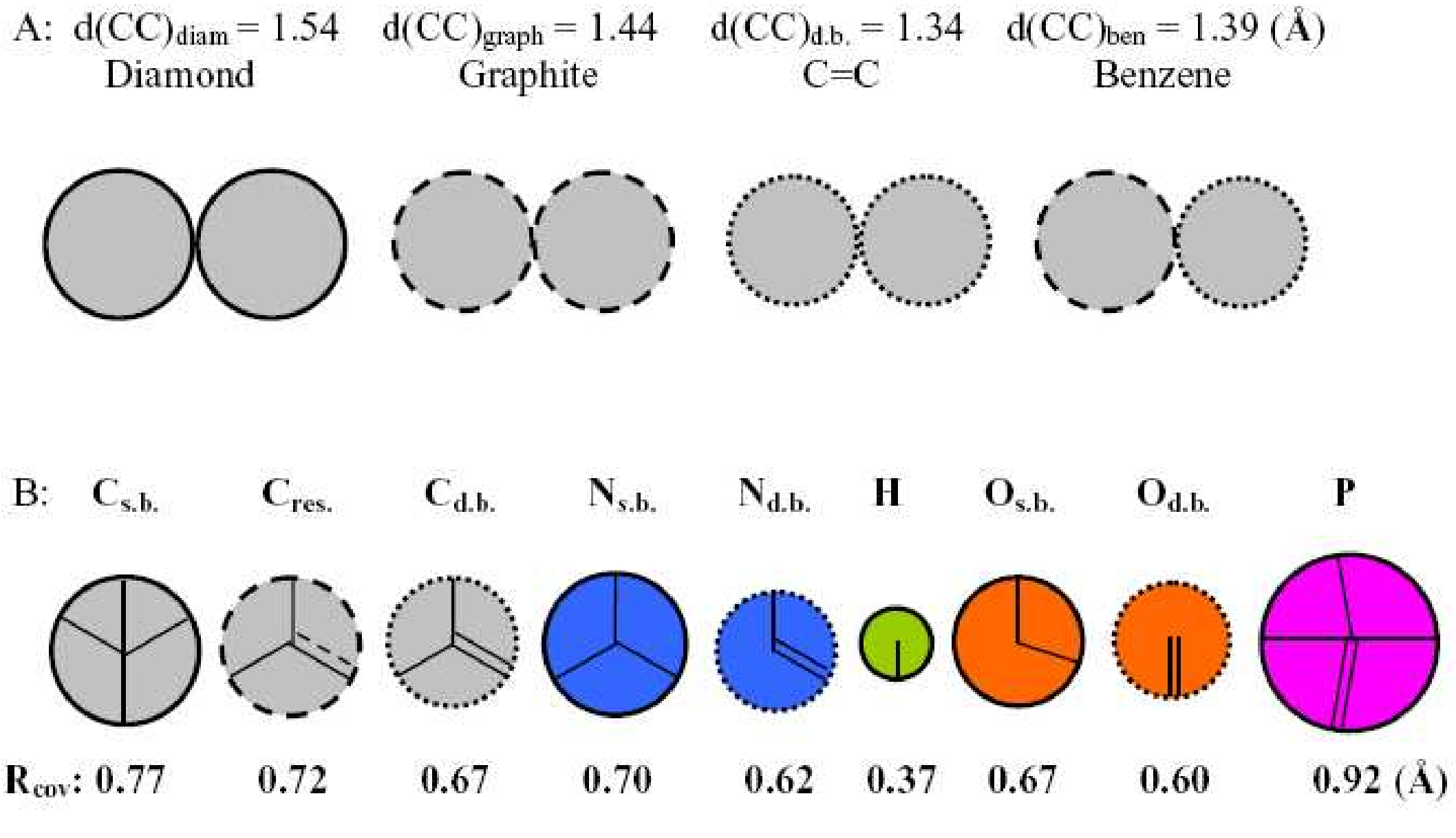

**Figure 2.** A: The covalent bond lengths d(CC) of four types of CC bonds. B: The (nine) covalent radii $R_{cov}$ = d(AA)/2 (+/- 0.02Å) [4] of the five atoms (mono to pentavalent: H, O, N, C, P) constituting the molecular components of nucleic acids. C: gray, N: blue, O: orange, H: green and P: pink. Nature makes unique combinations of these atoms to form the molecules of life (see Figs. 4 - 6).

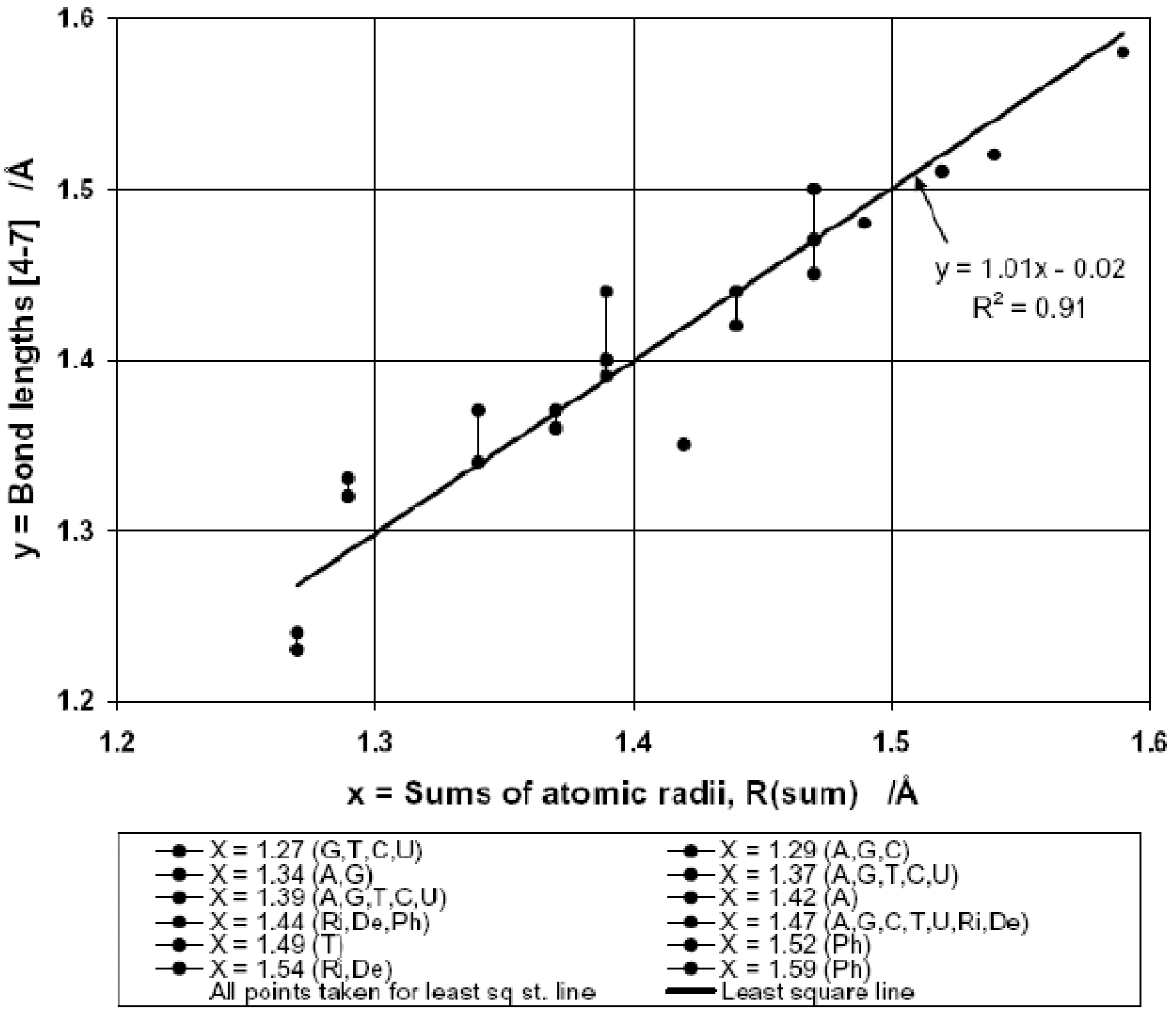

**Figure 3.** Graph of the bond lengths from [4-7] versus the sums of atomic covalent radii, R(sum), data in Table 3. The unit slope and zero intercept of the least squares line show that the bond lengths are equal to R(sum).

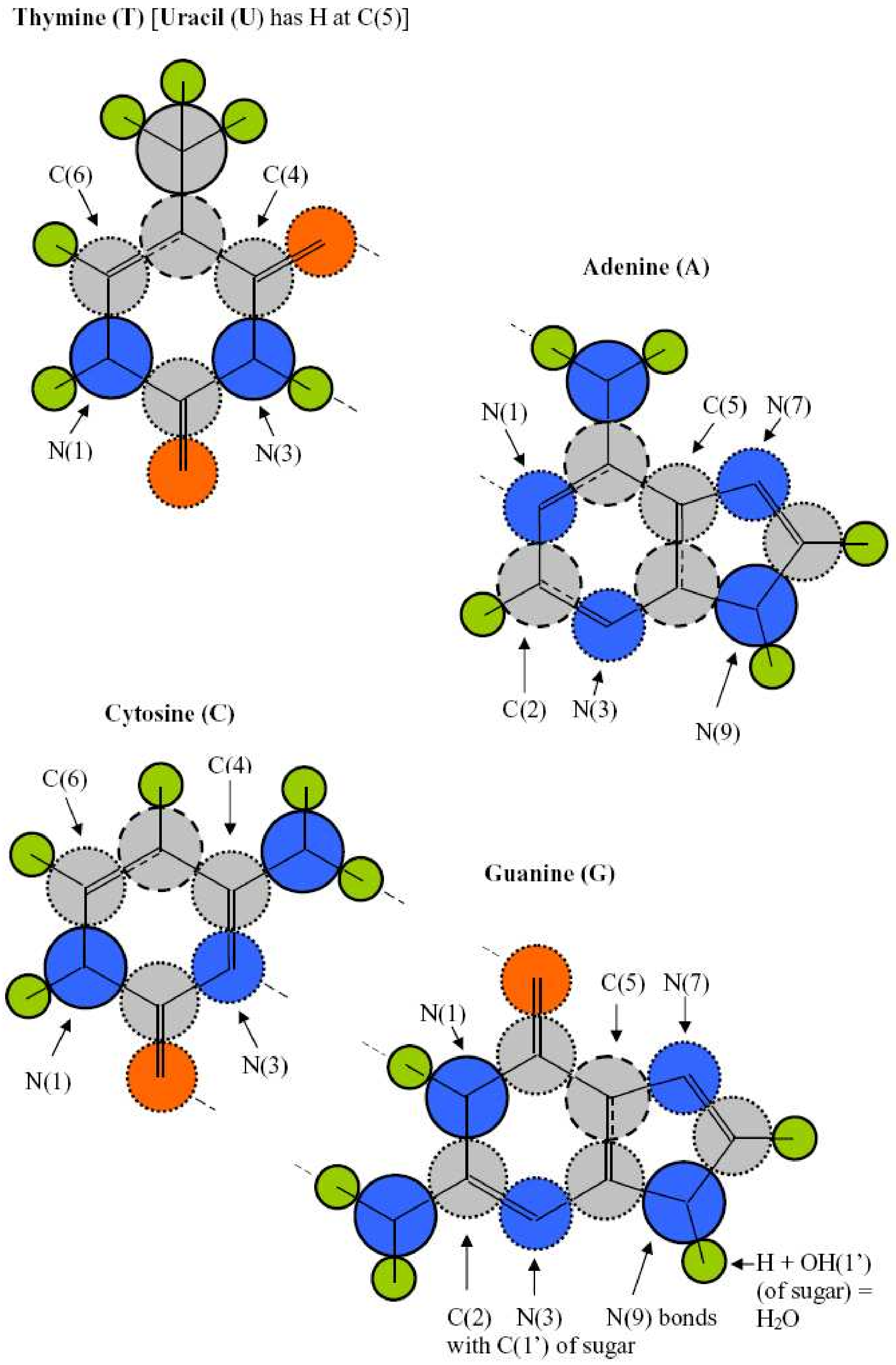

**Figure 4.** Atomic structures of the purines: A and G and pyrimidines: T, (U) and C. All bond lengths are sums of the covalent radii of the adjacent atoms (see Table 3). Base pairing [1a, b] occurs when hydrogen bonds connect T with A, and C with G along the broken lines, with bond lengths as in [3b].

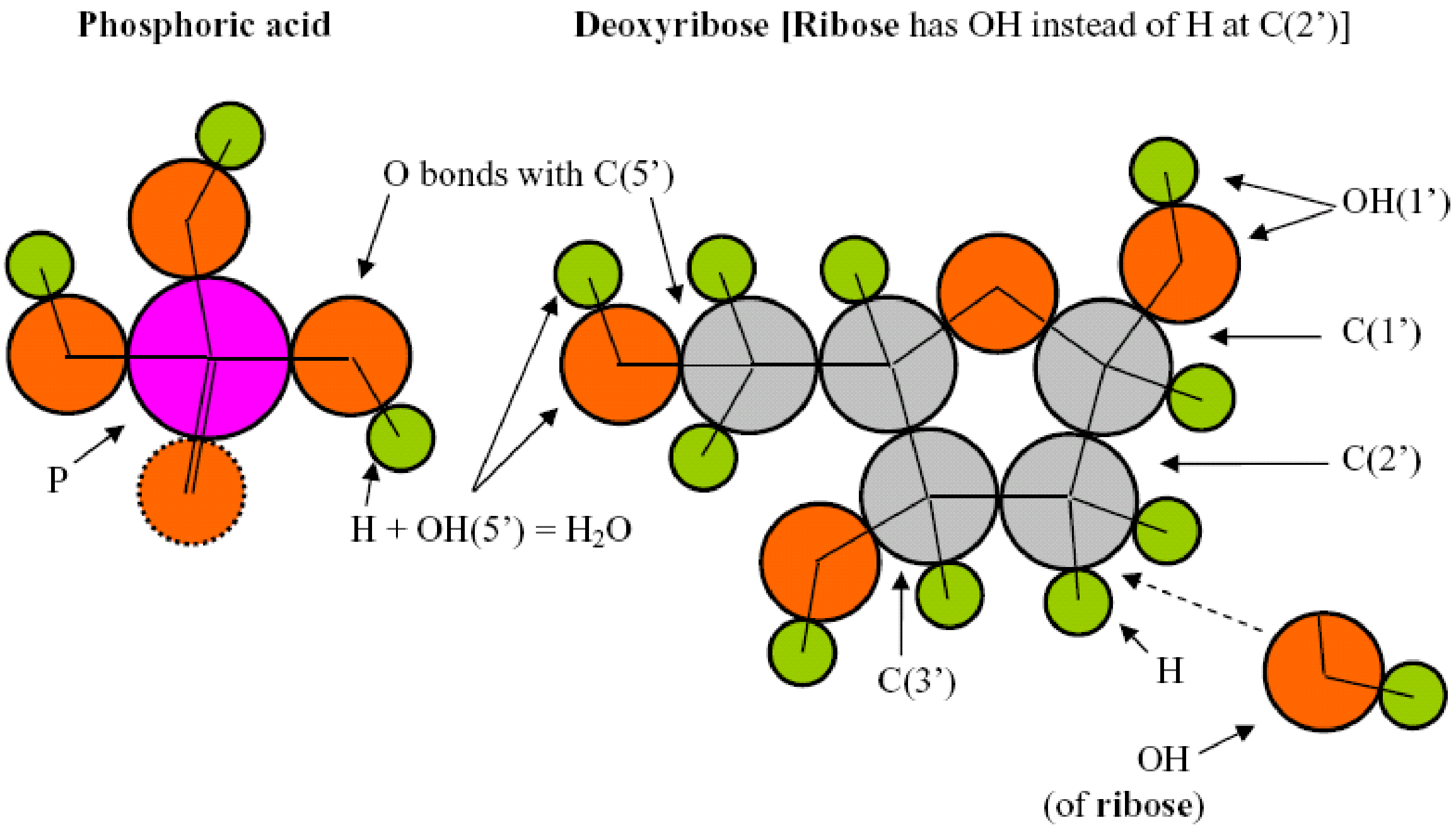

**Figure 5.** Atomic structures of phosphoric acid and deoxyribose (and ribose). All bond lengths are sums of the covalent radii of the adjacent atoms (see Table 3). By the elimination of water molecules as shown, the bases combine with sugars and the latter with phosphoric acid to form nucleotides [4-7].

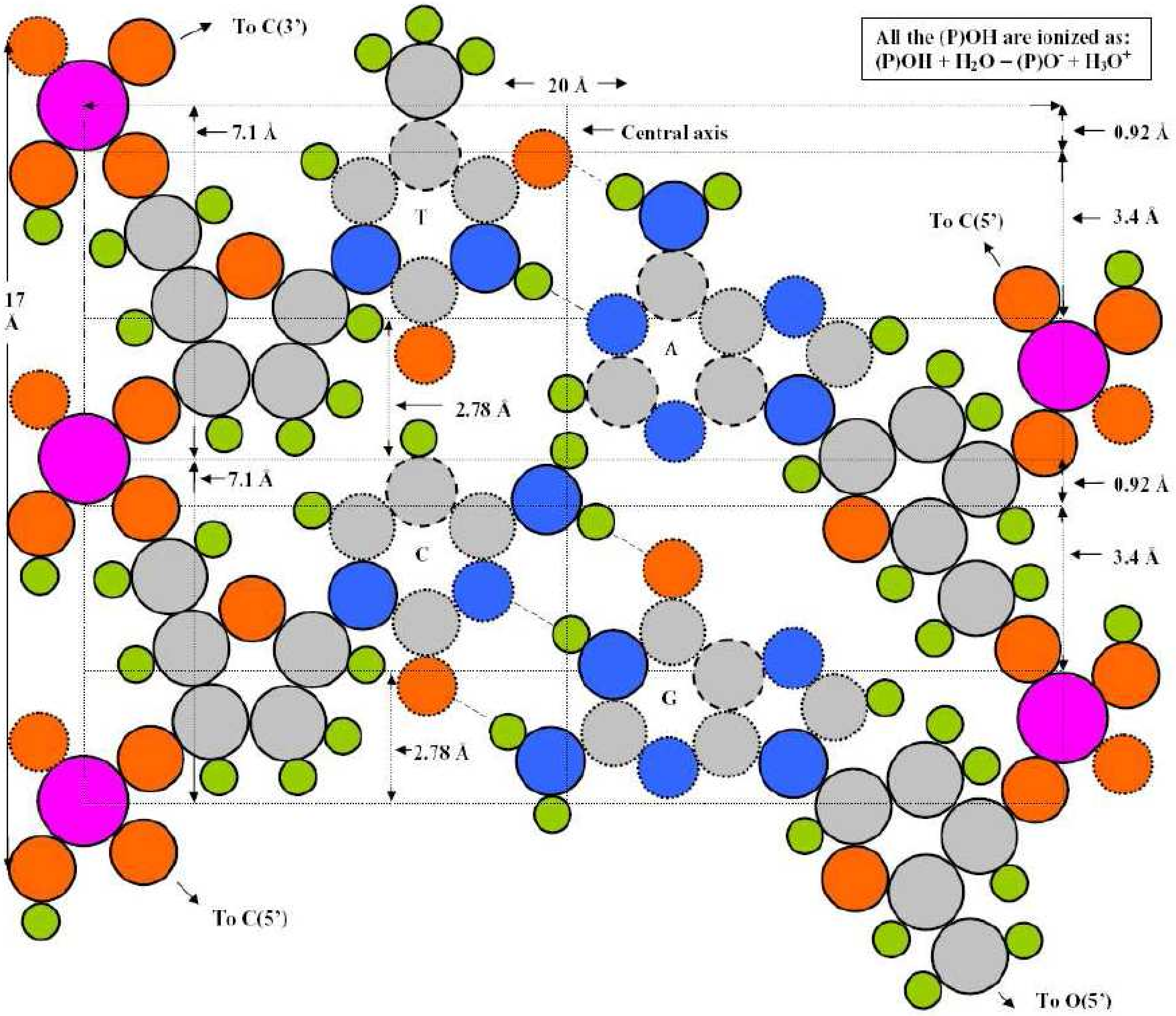

**Figure 6**. Atomic structures of the DNA nucleotides with the Watson-Crick base pairs (T, A and C, G connected by H-bonds as shown) attached to deoxyribose and phosphate molecules. As per the known crystallographic distances [1c], the two P atoms (pink) attached via O(5') and O(3') (orange) to the sugar molecule are 7.1 Å apart and the distance of each P atom from the central axis is 10 Å. The base pairs are perpendicular to the plane of the sugar-phosphate back bone and form the steps of the ladder of the double helix. The steps are 3.4 Å apart [1c]. The distance of 17 Å containing 5 P atoms is half of the 34 Å turn (with 10 P atoms) of the helix and the distance 2.78 Å can be considered as the thickness of the steps. (For the colors, radii and bonds of the atoms see Figs. 2, 4 and 5.)
Note: The above Figure holds also for RNA, but with Uracil (U) and Ribose in place of Thymine (T) and deoxyribose respectively.